\documentclass[twocolumn]{aastex631} 

\usepackage{mathtools}

\begin{document}

\title{An analytic model for the sub-galactic matter power spectrum in fuzzy dark matter halos
}

\author{Hiroki Kawai}
\affiliation{Department of Physics, University of Tokyo, Tokyo 113-0033, Japan}

\author[0000-0003-3484-399X]{Masamune Oguri}
\affiliation{Research Center for the Early Universe, University of Tokyo, Tokyo 113-0033, Japan}
\affiliation{Department of Physics, University of Tokyo, Tokyo 113-0033, Japan}
\affiliation{Kavli Institute for the Physics and Mathematics of the Universe (Kavli IPMU, WPI), University of Tokyo, Chiba 277-8582, Japan}

\author{Alfred Amruth}
\affiliation{Department of Physics, University of Hong Kong, Hong Kong, Hong Kong SAR}

\author{Tom Broadhurst}
\affiliation{Department of Theoretical Physics, University of the Basque Country, UPV/EHU, 48080 Bilbao, Spain}
\affiliation{Donostia International Physics Center (DIPC), 20018 Donostia, The Basque Country}
\affiliation{Ikerbasque, Basque Foundation for Science, E-48011 Bilbao, Spain}

\author{Jeremy Lim}
\affiliation{Department of Physics, University of Hong Kong, Hong Kong, Hong Kong SAR}

\begin{abstract}
Fuzzy dark matter (FDM), a scalar particle coupled to the gravitational field without self-interaction whose mass range is $m \sim 10^{-24} - 10^{-20}\ \rm{eV}$, is one of the promising alternative dark matter candidates to cold dark matter. 
The quantum interference pattern, which is a unique structure of FDM, can be seen in halos in cosmological FDM simulations.
In this paper, we first provide an analytic model of the sub-galactic matter power spectrum originating from  quantum clumps in FDM halos, in which the density distribution of the FDM is expressed by a superposition of quantum clumps whose size corresponds to the de Broglie wavelength of the FDM. 
These clumps are assumed to be distributed randomly such that the ensemble averaged density follows the halo profile such as the Navarro-Frenk-White profile. 
We then compare the convergence power spectrum projected along the line of sight around the Einstein radius, which is converted from the sub-galactic matter power spectrum, to that measured in the strong lens system SDSS J0252+0039.
While we find that the current observation provides no useful constraint on the FDM mass, we show that future deep, high spatial resolution observations of strong lens systems can tightly constrain FDM with the mass around $10^{-22}\ \rm{eV}$.
\end{abstract}

\keywords{Dark matter (353) --- Strong gravitational lensing (1643) --- Galaxy dark matter halos (1880)}

\section{Introduction} \label{sec:intro}
Dark matter is one of the major components in the Universe, yet its nature is not fully understood. 
In the standard $\Lambda$ dominated cold dark matter ($\Lambda$CDM) cosmology, the energy density of the Universe is composed of baryon ($\sim 5\%$), cold dark matter ($\sim 25\%$), and dark energy ($\sim 70\%$), according to the cosmic microwave background (CMB) observation \citep{2020A&A...641A...6P}
The $\Lambda$CDM model can successfully explain the large scale structure of the Universe.
However, there are some discrepancies below the scale of $\sim 1 \ \rm Mpc$ between CDM predictions and observations, which are often referred to as small scale problems (e.g. \citealp{2017Galax...5...17D, 2017ARA&A..55..343B} for review). 
These include the core-cusp problem (e.g. \citealt{2001AJ....122.2381M}), the diversity problem \citep{2015MNRAS.452.3650O}, the missing satellite problem (e.g. \citealt{1999ApJ...522...82K}), and the too-big-to-fail problem (e.g. \citealt{2011MNRAS.415L..40B}). 
It is not yet clear whether these discrepancies on small scales in the $\Lambda$CDM Universe originate from baryonic feedback such as supernova explosion or the unknown nature of dark matter, or even both. 
While the small scale crisis might be resolved by including baryonic process, the possibility of resolving the crisis by changing the nature of dark matter has also been extensively studied.

Fuzzy dark matter \citep[FDM,][]{2000PhRvL..85.1158H} is one of the promising dark matter candidates that might resolve both the core-cusp problem and the missing satellite problem. 
It is a scalar particle coupled only to gravity. The typical mass is around $10^{-22}\ \rm{eV}$. Such a small mass results in its de Broglie wavelength to be $\mathcal{O}(1)\ \rm{kpc}$ that is an important scale for the small scale problems. 
At the scale below the de Broglie wavelength, the wavelike nature can be seen, while at the larger scale FDM behaves similarly to CDM. 
FDM simulations reveal nature of FDM halos \citep{2014NatPh..10..496S}.
The inner region consists of a core whose size is about the de Broglie wavelength of FDM, while the outer region follows the Navarro-Frenk-White (NFW) profile \citep{1997ApJ...490..493N} in the similar way as in CDM halos.
In the FDM halo, granular structures can also be seen, which is a distinctive feature caused by the wavelike nature of FDM.
Throughout the paper we call them quantum clumps.

Currently there are several constraints on the FDM mass range (\citealt{2020arXiv200503254F} for review). 
Constraints are obtained from CMB power spectrum \citep{2015PhRvD..91j3512H,2018MNRAS.476.3063H} and Lyman alpha forests \citep{2019MNRAS.482.3227N, 2017MNRAS.471.4606A, 2017PhRvL.119c1302I, 2021PhRvL.126g1302R}, which provide the most stringent constraint.
These are the constraints from observations at the scale of $k \geq 10 \rm Mpc^{-1}$, suggesting that they do not directly come from the small scale.
Future observations of 21 cm line from neutral hydrogen can reach below that scale  \citep{2018PhRvD..98b3011L, 2018PhRvD..98f3021S, 2021ApJ...913....7J} and can tightly constrain FDM.
There are many other constraints from various observations, such as the subhalo mass function \citep{2020PhRvD.101l3026S, 2020PhRvD.101j3023B}, the dynamical friction \citep{2020JCAP...01..001L}, the Milky Way \citep{2019MNRAS.485.2861C}, dwarf spheroidals, \citep{2017MNRAS.468.1338C, 2017MNRAS.472.1346G}, dwarf satellites \citep{2020ApJ...893...21S}, Ultra-faint dwarf spheroidals \citep{2021ApJ...912L...3H}, and rotation curves \citep{2020PhRvD.101j3504M}.
Some of these constraints exclude the most interesting FDM mass range around  $10^{-22}\ \rm{eV}$.
However since these constraints suffer from systematic uncertainties mainly originating from baryon physics, it is important to derive constraints by several different approaches to cross check possible systematic effects.

Recently a new framework to study sub-galactic structure has been proposed.
\citet{2016JCAP...11..048H} constructed the framework to estimate the projected matter power spectrum by using strong gravitational lensing. 
The method to estimate sub-galactic convergence power spectrum from substructures in lens plane is studied in \citet{2018PhRvD..97b3001D}. 
\citet{2018arXiv180305952B} applied these methods to the real strong lens system SDSS J0252+0039 that is a galaxy-galaxy strong lens system found in SLACS survey \citep{2009ApJ...705.1099A}.
They obtained constraints on the sub-galactic matter power spectrum and discussed the implication for the CDM model.
We expect that this method can also be applied to the FDM model.
Since a FDM halo is filled with quantum clumps, the FDM model should predict the much higher amplitude of the sub-galactic matter power spectrum on the scale of the size of the quantum clumps.

In this paper, we first construct an analytic model for the sub-galactic matter power spectrum in FDM halos.
We assume that the FDM halos consist of quantum clumps whose size is fixed to the de Broglie wavelength of FDM.
These clumps are assumed to be distributed randomly such that the ensemble average of the FDM density reduces to the assumed halo profile. 
We also include baryon components that are assumed to be independent of the FDM component. 
We then calculate the sub-galactic matter power spectrum under these assumptions. 
Finally, we compare our model with the constraints from SDSS J0252+0039. 
Note that \citet{2020PhRvL.125k1102C} pointed out that multiple images and flux ratio anomalies in strong lens systems can be observed more often if the halo is composed of FDM. 
Our work is complementary to their work in that we focus on the sub-galactic power spectrum rather than flux ratio anomalies.

This paper is organized as follows. In Sec.~\ref{sec:power-spectrum}, we present our model to calculate the sub-galactic matter power spectrum in FDM halos. 
We show the dependence of parameters on the sub-galactic matter power spectrum.
In Sec.~\ref{sec:compare-obs}, we compare our model with current observational data obtained by a  strong gravitational lens system. 
In addition, we show that future observations can constrain the interesting range of the FDM mass. Finally we conclude in Sec.~\ref{sec:conclusion}.

\section{Sub-galactic matter power spectrum of FDM halo} \label{sec:power-spectrum}
The property of FDM halos is studied by FDM simulations (e.g. \citealt{2014NatPh..10..496S}). 
The simulations reveal a pervasive modulation of the density field on the scale of de Broglie wavelength, which can naturally be approximated by FDM halos being filled with quantum clumps whose size is roughly the de Broglie wavelength of FDM.
These clumps produce large matter fluctuations on the scale of the de Broglie wavelength.
In order to evaluate these fluctuations, we construct an analytic model of the sub-galactic matter power spectrum in FDM halos including baryon. 
In our model, FDM halos are assumed to be consist of quantum clumps.
They are distributed randomly, while the ensemble averaged density reduces to a specific dark matter profile (e.g. NFW profile, \citealt{1997ApJ...490..493N}).
The density profile of baryon is assumed to be independent of FDM and to follow a specific baryon profile (e.g. Hernquist profile, \citealt{1990ApJ...356..359H}).
In Sec.~\ref{subsec:fdmonly}, we give a formulation in the FDM-only case.
We then include the baryon profile in Sec.~\ref{subsec:with-baryon}.
In Sec.~\ref{subsec:parameter-dep}, we show the result when the dark matter and baryon profiles follow NFW and Hernquist profiles, respectively, and discuss the parameter dependence of the power spectrum.

\subsection{FDM-only case} \label{subsec:fdmonly}
The distribution of FDM in a halo is determined by that of qunatum clumps.
We assume that the mass of each clump $M_{\rm c}$ at $\boldsymbol{r'}$ is determined from an average local density of the halo $\rho_{\rm h}(\boldsymbol{r'})$ as
\begin{equation}
    M_{\rm c}(\boldsymbol{r'}) = \rho_{\rm h}(\boldsymbol{r'})V_{\rm c} \label{clump_mass},
\end{equation}
where
\begin{equation}
    V_{\rm c} = \frac{4}{3} \pi \left(\frac{\lambda_{\rm c}}{2}\right)^{3}
\end{equation}
is the volume of each clump whose radius is assumed to be given by the half of the de Broglie wavelength $\lambda_{\rm c} = 2 \pi \hbar/mv$ with $m$ being the FDM mass and $v$ being its velocity dispersion.
Here we assume that $v$ can be approximated as a constant value within a halo, which indicates that $V_{\rm c}$ is constant in our formalism.

Inside each clump, the density profile $\rho_{\rm c}(\boldsymbol{r})$ can be described using the normalized mass profile function $u(\boldsymbol{r}-\boldsymbol{r'})$ around quantum clump whose center is located at $\boldsymbol{r'}$ as 
\begin{equation}
    \rho_{\rm c}(\boldsymbol{r}\ ;\boldsymbol{r'}) = M_{\rm c}(\boldsymbol{r'}) u(\boldsymbol{r}-\boldsymbol{r'}).
\end{equation}
The normalization condition is
\begin{equation}
    \int_{V} d^{3}r\ u(\boldsymbol{r}-\boldsymbol{r'}) = \int_{V_{\epsilon}(\boldsymbol{r'})} d^{3}r\ u(\boldsymbol{r}-\boldsymbol{r'}) = 1, \label{normalize}
\end{equation}
where $V$ is the total volume of the halo and $V_{\epsilon}(\boldsymbol{r'})$ is a three dimensional sphere around the point $\boldsymbol{r'}$ that is sufficiently small compared with the size of the halo but is larger than the size of each clump.
As discussed later, in this paper we assume a smoothly truncated normalized mass density profile such that its tail extends much beyond the de Broglie wavelength. 
This assumption leads to overlaps of FDM mass densities among neighboring clumps.

We assume that these quantum clumps are randomly distributed on the small scale, while the ensemble average of the number density is fixed.
Since we assume that the mass of these clumps depends on the location (Eq.~\ref{clump_mass}), the ensemble average of randomly distributed clumps can yield non-uniform density distributions within a halo such as the NFW density profile, as we discuss in more detail later.  
The density profile of the FDM halo $\rho_{\rm f}(r)$ can be expressed by a superposition of these randomly distributed clumps. 
Under this assumption, the FDM halo profile is given by
\begin{eqnarray}
    \rho_{\rm f}(\boldsymbol{r}) &=& \int_{V} d^{3}r' \rho_{\rm c}(\boldsymbol{r}\ ;\boldsymbol{r'}) n(\boldsymbol{r'}) \nonumber \\
    &=& \int_{V} d^{3}r' \rho_{\rm h}(\boldsymbol{r'}) V_{\rm c} n(\boldsymbol{r'}) u(\boldsymbol{r}-\boldsymbol{r'}), \label{rho_f}
\end{eqnarray}
where $n(\boldsymbol{r})$ represents the number density of the center of quantum clumps.
Suppose each clump is indexed by $j$ and its center by $r_{j}$, we can rewrite $n(\boldsymbol{r'})$ in terms of the Dirac delta function,
\begin{equation}
    n(\boldsymbol{r'}) = \sum_{j} \delta^{(3)}(\boldsymbol{r'}-\boldsymbol{r'}_{j}).
\end{equation}
Note that $\rho_{\rm f}(\boldsymbol{r})$ takes a constant value inside $V_{\epsilon}$ sphere since the sphere is assumed to be sufficient small. We use this assumption in the following calculation.
The ensemble average of the number density $\left<n(\boldsymbol{r})\right>$ of quantum clumps is set to $1/V_{\rm c}$ such that the ensemble average of the FDM density $\left<\rho_{\rm f}(\boldsymbol{r})\right>$ reduces to the assumed average halo profile,
\begin{equation}
    \left<\rho_{\rm f}(\boldsymbol{r})\right> = \rho_{\rm h}(\boldsymbol{r}).
\end{equation}
Note also that we do not model the soliton core specifically since we focus on the strong lens region which is larger than the typical radius of the soliton core.
More detailed discussions on the validity of our analytic model are given in Sec.~\ref{subsec:parameter-dep}.

Using the three dimensional density field of the FDM halo derived above, we define the two dimensional density field $\Sigma_{\rm f}(\boldsymbol{x})$ projected along the line of sight that is aligned with $z$ axis,
\begin{eqnarray}
    \Sigma_{\rm f}(\boldsymbol{x}) &\equiv& \int_{Z} dz\ \rho_{\rm f}(\boldsymbol{r}) \label{sigma_def} \nonumber \\
    &=& \int_{Z} dz \int_{V_{\epsilon(\boldsymbol{r})}} d^{3}r'\  \rho_{\rm h}(\boldsymbol{r'}) V_{\rm c} n(r') u(\boldsymbol{r}-\boldsymbol{r'}),
\end{eqnarray}
where $\boldsymbol{x}$ is the position in the projected two dimensional coordinates and $Z$ represents the range of the integration along the line of sight.
The second line can be obtained by using the normalization condition Eq.~(\ref{normalize}).
The ensemble average of projected density field $\left<\Sigma_{\rm f}(\boldsymbol{x})\right>$ is given by
\begin{eqnarray}
    \left<\Sigma_{\rm f}(\boldsymbol{x})\right> &=& \int_{Z} dz \left<\rho_{\rm f}(\boldsymbol{r})\right> = \int_{Z} dz\ \rho_{\rm h}(\boldsymbol{r}) \nonumber \\
    &\equiv& \Sigma_{\rm h}(\boldsymbol{x}).
\end{eqnarray}

In order to calculate the sub-galactic matter power spectrum, we consider a sufficiently small two dimensional sphere $S_{\epsilon}$ in the projected density field around $\boldsymbol{x}$. 
In the same way as in the three dimensional case, we assume that $\Sigma_{\rm h}(\boldsymbol{x})$ is constant inside $S_{\epsilon}$ sphere.
Eq.~(\ref{sigma_def}) can be rewritten by using $S_{\epsilon}$ and Eq.~(\ref{rho_f}) as
\begin{equation}
    \Sigma_{\rm f}(\boldsymbol{x}) = \int_{Z} dz \int_{S_{\epsilon} \times Z} d^{3}r' \rho_{\rm h}(\boldsymbol{r'}) V_{\rm c} n(\boldsymbol{r'}) u(\boldsymbol{r}-\boldsymbol{r'}),
\end{equation}
where the integral range $S_{\epsilon} \times Z$ represents the cylinder along the line of sight. 

The density fluctuation is given by
\begin{eqnarray}
    \delta(\boldsymbol{x}) &\equiv& \frac{\Sigma_{\rm f}(\boldsymbol{x}) - {\Sigma_{\rm h}}(\boldsymbol{x})}{\Sigma_{\rm h}(\boldsymbol{x})} \nonumber \\
    &=& \frac{1}{\Sigma_{\rm h}(\boldsymbol{x})} \int_{Z} dz \int_{S_{\epsilon} \times Z} d^{3}r' \rho_{\rm h}(\boldsymbol{r'}) V_{\rm c} n(\boldsymbol{r'}) u(\boldsymbol{r}-\boldsymbol{r'}) \nonumber \\
    &&\hspace{5.0cm} - 1. \label{fluctuation}
\end{eqnarray}
The Fourier transform of the density fluctuation is 
\begin{eqnarray}
    \widetilde{\delta}_{\boldsymbol{k}}\ &\equiv& \int_{S_{\epsilon}} d^{2}x\ \delta(\boldsymbol{x}) e^{-i\boldsymbol{k} \cdot \boldsymbol{x}} \nonumber \\
    &=& \frac{1}{\Sigma_{\rm h}(\boldsymbol{x})} \int_{S_{\epsilon} \times Z} d^{3}r' \rho_{\rm h}(\boldsymbol{r'}) V_{c} n(r') \nonumber \\
    &{}&\hspace{1cm} \times \int_{S_{\epsilon} \times Z} d^{3}r\ u(\boldsymbol{r}-\boldsymbol{r'}) e^{-i\boldsymbol{K} \cdot \boldsymbol{r}}\mid_{K_{z}=0} \nonumber \\
    &=& \frac{V_{c}}{\Sigma_{\rm h}(\boldsymbol{x})} \widetilde{u}_{\boldsymbol{k}} \int_{S_{\epsilon} \times Z} d^{3}r' \rho_{\rm h}(\boldsymbol{r'}) \nonumber \\
    &{}&\hspace{2.5cm} \times n(r') e^{-i\boldsymbol{K} \cdot \boldsymbol{r'}}\mid_{k_{z}=0}, \label{fl_fourier}
\end{eqnarray}
where $\boldsymbol{k}$ represents the two dimensional wavenumber of the fluctuation and $\boldsymbol{K}$ is the three dimensional wavenumber that is related with $\boldsymbol{k}$ as $K_{x}=k_{x}, K_{y}=k_{y}, K_{z}=0$.
Note that we ignore the last term in Eq.~(\ref{fluctuation}) since we are not interested in the fluctuation with $\boldsymbol{k} = \boldsymbol{0}$.
We also assume that $Z$ is sufficiently large compared with the extent of the normalized mass profile function $u(\boldsymbol{r})$.
The Fourier transform of this function is denoted as $\widetilde{u}_{\boldsymbol{K}}$, and is related with $\widetilde{u}_{\boldsymbol{k}}$ as 
\begin{equation}
    \widetilde{u}_{\boldsymbol{k}} \equiv \widetilde{u}_{\boldsymbol{K}}\mid_{K_{z}=0}.
\end{equation}

The definition of the sub-galactic matter power spectrum is
\begin{equation}
    \left<\widetilde{\delta}_{\boldsymbol{k}}\ \widetilde{\delta}_{\boldsymbol{k'}} \right> \equiv S_{\epsilon} \delta^{(2)}_{\boldsymbol{k}+\boldsymbol{k'}, \boldsymbol{0}}\ P(k). \label{ps_def}
\end{equation}
The left hand side of Eq.~(\ref{ps_def}) can be calculated by substituting Eq.~(\ref{fl_fourier}) as
\begin{eqnarray}
    &{}&\left<\widetilde{\delta}_{\boldsymbol{k}}\ \widetilde{\delta}_{\boldsymbol{k'}} \right> \nonumber \\
    &=& \left(\frac{V_{\rm c}}{\Sigma_{\rm h}(\boldsymbol{x})}\right)^{2} \widetilde{u}_{\boldsymbol{k}}\ \widetilde{u}_{\boldsymbol{k'}} \int_{S_{\epsilon} \times Z} d^{3}r \int_{S_{\epsilon} \times Z} d^{3}r' \left< n(\boldsymbol{r}) n(\boldsymbol{r'})\right> \nonumber \\ &{}&\hspace{1cm} \times \rho_{\rm h}(\boldsymbol{r}) \rho_{\rm h}(\boldsymbol{r'}) e^{-i\boldsymbol{K} \cdot \boldsymbol{r}} e^{-i\boldsymbol{K'} \cdot \boldsymbol{r'}}\mid_{K_{z}=0, K'_{z}=0} \nonumber \\
    &=& \left(\frac{V_{\rm c}}{\Sigma_{\rm h}(\boldsymbol{x})}\right)^{2} \widetilde{u}_{\boldsymbol{k}}\ \widetilde{u}_{\boldsymbol{k'}} \int_{S_{\epsilon} \times Z} d^{3}r\ \rho_{\rm h}^{2}(\boldsymbol{r}) \nonumber \\ &{}&\hspace{3cm} \times e^{-i(\boldsymbol{K}+\boldsymbol{K'}) \cdot \boldsymbol{r}}\mid_{K_{z}=0, K'_{z}=0} \nonumber \\
    &=& \left(\frac{V_{\rm c}}{\Sigma_{\rm h}(\boldsymbol{x})}\right)^{2} \widetilde{u}_{\boldsymbol{k}}\ \widetilde{u}_{\boldsymbol{k'}} \int_{Z} dz\ \rho_{\rm h}^{2}(\boldsymbol{r}) \int_{S_{\epsilon}} d^{2}x\  e^{-i(\boldsymbol{k}+\boldsymbol{k'}) \cdot \boldsymbol{x}} \nonumber \\
    &=& S_{\epsilon} \delta^{(2)}_{\boldsymbol{k}+\boldsymbol{k'}, \boldsymbol{0}}\ \frac{V_{\rm c}}{\Sigma_{\rm h}^{2}(\boldsymbol{x})}\ |\widetilde{u}_{\boldsymbol{k}}|^{2} \int_{Z} dz\ \rho_{\rm h}^{2}(\boldsymbol{r}). \label{lhs}
\end{eqnarray}
In the second equality, we assume there is no correlation between the number density at different positions, $\left< n(\boldsymbol{r}) n(\boldsymbol{r'})\right> = \delta^{(3)}(\boldsymbol{r}-\boldsymbol{r'})$. 
From Eq.~(\ref{ps_def}) and Eq.~(\ref{lhs}), we can finally obtain the sub-galactic matter power spectrum in FDM halos,
\begin{equation}
    P_{\rm f}(k) = \frac{V_{\rm c}}{r_{\rm h}(\boldsymbol{x})}\ |\widetilde{u}_{\boldsymbol{k}}|^{2}, \label{ps_vec}
\end{equation}
where the effective halo size $r_{\rm h}(\boldsymbol{x})$ has an unit of the length and is given by
\begin{equation}
    r_{\rm h}(\boldsymbol{x}) \equiv \frac{\Sigma_{\rm h}^{2}(\boldsymbol{x})}{\int_{Z} dz\ \rho_{\rm h}^{2}(\boldsymbol{r})} = \frac{\left(\int_{Z} dz\ \rho_{\rm h}(\boldsymbol{r})\right)^{2}}{\int_{Z} dz\ \rho_{\rm h}^{2}(\boldsymbol{r})}. \label{rh_vec}
\end{equation}
Here we add a subscript f on $P(k)$ since we only consider the FDM component.
Assuming spherically symmetric halos and matter profile function of each clump, Eq.~(\ref{ps_vec}) can be further simplified as
\begin{equation}
    P_{\rm f}(k) = \frac{V_{\rm c}}{r_{\rm h}(x)}\ |\widetilde{u}_{k}|^{2}, \label{ps}
\end{equation}
\begin{equation}
    r_{\rm h}(x) = \frac{\Sigma_{\rm h}^{2}(x)}{\int_{Z} dz\ \rho_{\rm h}^{2}(r)} = \frac{\left(\int_{Z} dz\ \rho_{\rm h}(r)\right)^{2}}{\int_{Z} dz\ \rho_{\rm h}^{2}(r)}. \label{rh}
\end{equation}
While clumps in FDM simulations appear to be non-spherical (e.g. \citealt{2014NatPh..10..496S}), we assume them to be spherical for simplicity. The validity of this assumption in our model is discussed in Sec.~\ref{subsec:parameter-dep}.

Here we show that the effective halo size $r_{\rm h}(x)$ contains the information of the density dispersion along the line of sight. We can rewrite Eq.~(\ref{rh}) as
\begin{eqnarray}
    r_{\rm h}(x) &=& Z\cdot \frac{\left(\frac{1}{Z} \int_{Z} dz\ \rho_{\rm h}(r)\right)^{2}}{\frac{1}{Z} \int_{Z} dz\ \rho_{\rm h}^{2}(r)} \nonumber\\
    &=& Z\cdot \frac{\overline{\rho}_{\rm h}^{2}(x)}{\overline{\rho}_{\rm h}^{2}(x) + s^{2}(x)}, \label{rh_var}
\end{eqnarray}
where $\overline{\rho}_{\rm h}(x)$ represents the average halo density along the line of sight and $s^{2}(x)$ represents the density dispersion, which are given by
\begin{equation}
    \overline{\rho}_{\rm h}(x) \equiv \frac{1}{Z} \int_{Z} dz\ \rho_{\rm h}(r),
\end{equation}
\begin{equation}
    s^{2}(x) \equiv \frac{1}{Z} \int_{Z} dz\ \rho_{\rm h}^{2}(r) - \left(\frac{1}{Z} \int_{Z} dz\ \rho_{\rm h}(r)\right)^{2}. \label{s}
\end{equation}

\begin{figure}
 \centering
 \includegraphics[width=85mm]{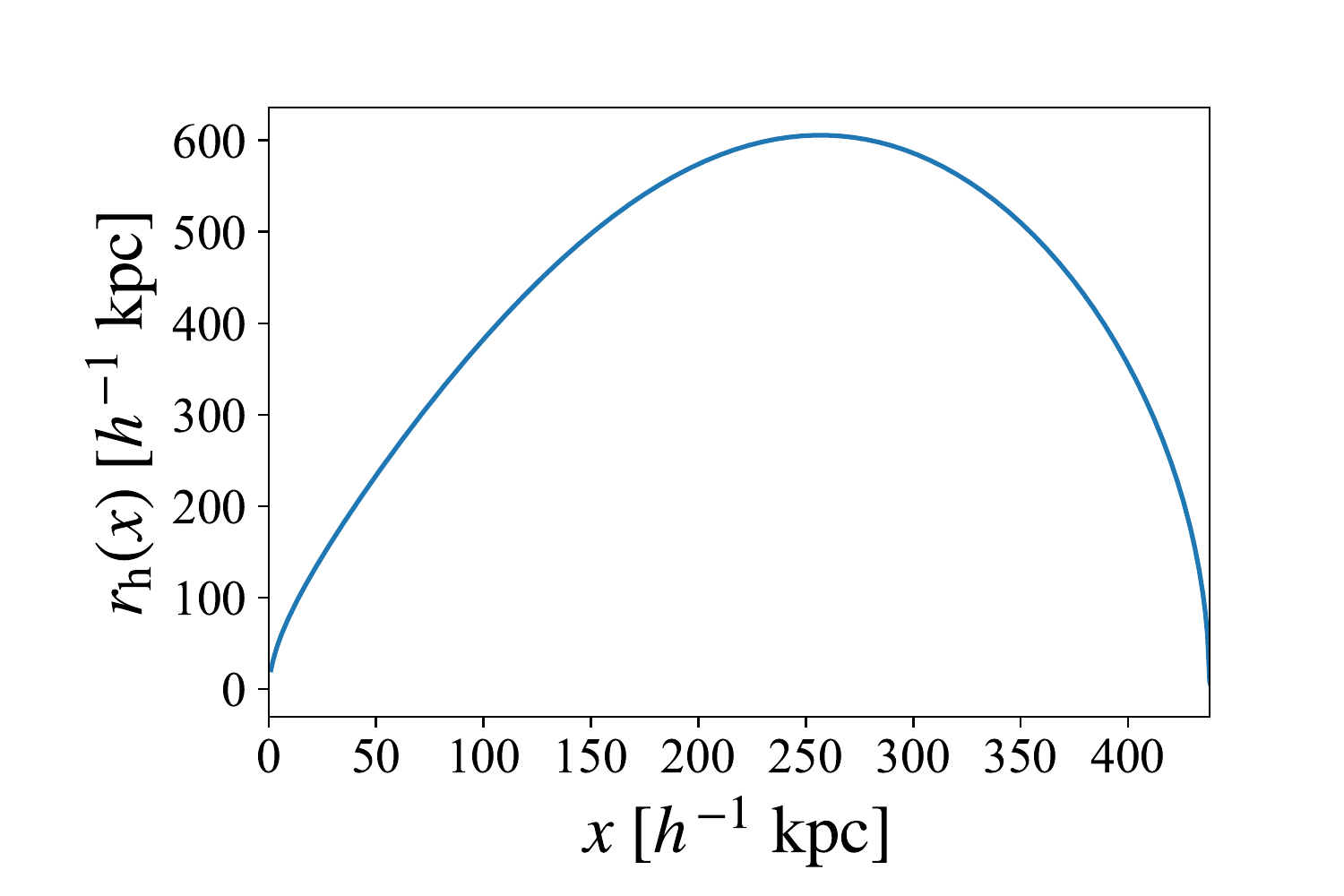}
 \caption{The effective halo size $r_{\rm h}(x)$ as a function of radius $x$ in the projected density field. We use the NFW profile as a halo profile. The total halo mass is set to $M_{\rm h} = 10^{13}\ h^{-1}\ M_{\odot}$ and we use a relation between the concentration parameter and the total mass of \citet{2021MNRAS.tmp.1529I}. The integration along the line of sight is limited to the virial radius, about $438\ h^{-1}\ \rm kpc$.} 
 \label{rh_fig}
\end{figure}

Fig.~\ref{rh_fig} shows an example of $r_{\rm h}(x)$ assuming an NFW profile as the halo density profile.
It is seen that $r_{\rm h}(x)$ is monotonically increasing around the central region, while it is monotonically decreasing in the outer region.
This behavior can be understood from Eq.~(\ref{rh_var}). 
Around the central region, $s(x)$ determines the increase/decrease of $r_{\rm h}(x)$ since the density dispersion along the line of sight is large.
In the outer region, the halo size along the line of sight $Z$ determines the shape of $r_{\rm h}(x)$.

\subsection{Including baryon} \label{subsec:with-baryon}
In Sec.~\ref{subsec:fdmonly}, we describe the sub-galactic matter power spectrum of FDM-only halos. 
Since most of halos contain baryon, we also need to consider a baryon profile.
We assume that baryon is smoothly distributed with the smooth density profile function $\rho_{\rm b}(r)$. 
The total density $\rho(r)$ is
\begin{equation}
    \rho(r) = \rho_{\rm f}(r) + \rho_{\rm b}(r).
\end{equation}
The total projected density $\Sigma(x)$ is
\begin{equation}
    \Sigma(x) = \Sigma_{\rm f}(x) + \Sigma_{\rm b}(x), \label{Sigma_tot}
\end{equation}
where $\Sigma_{\rm b}(x)$ is defined as
\begin{equation}
    \Sigma_{\rm b}(x) \equiv \int_{Z} dz\ \rho_{\rm b}(r).
\end{equation}
Since we assume that baryon component does not contain any random component, the ensemble averaging of the baryon functions does not change their functional form.
We repeat the calculation in Sec.~\ref{subsec:fdmonly} to obtain the sub-galactic matter power spectrum with baryon,
\begin{equation}
    P(k) = \left(\frac{\Sigma_{\rm h}(x)}{\Sigma_{\rm h}(x) + \Sigma_{\rm b}(x)}\right)^{2} P_{\rm f}(k). \label{ps_wb}
\end{equation}
Eq.~(\ref{ps_wb}) indicates that power spectrum with baryon is smaller than that without baryon because the additional contribution of the smooth baryon component smears out the density fluctuations due to FDM.

\subsection{Parameter dependence} \label{subsec:parameter-dep}
We calculate the power spectrum Eq.~(\ref{ps_wb}) with specific functions.
Halo and baryon profiles are set to the NFW and the Hernquist profiles, respectively.
While the NFW profile has two parameters, the total halo mass and the concentration parameter, it is known that there is a scaling relation between them (e.g. \citealt{2021MNRAS.tmp.1529I}).
Assuming that relation, we need only one parameter, the total halo mass denoted by $M_{\rm h}$.
Calculations are conducted using the python module {\sc Colossus} \citep{2018ApJS..239...35D}.
The Hernquist profile has two parameters, the total stellar mass and the characteristic radius.
The empirical relation between them is also known by fitting a sample of 50000 early-type galaxies \citep{2009MNRAS.394.1978H}.
We thus use the single parameter $M_{\rm s}$ to determine the Hernquist profile.
Note that we use the stellar-to-halo mass ratio $M_{\rm s}/M_{\rm h}$ as a parameter instead of the stellar mass $M_{\rm s}$. 
For sufficiently high mass halos with masses larger than about $10^{11}\ M_{\odot}$, the stellar-to-halo mass ratio in the FDM model is expected to be the same as in the CDM model \citep{2019MNRAS.482.4364C} and is known to be around $10^{-3} - 10^{-1}$ \citep{2018ARA&A..56..435W} in this halo mass range.

\begin{figure}
 \centering
 \includegraphics[width=85mm]{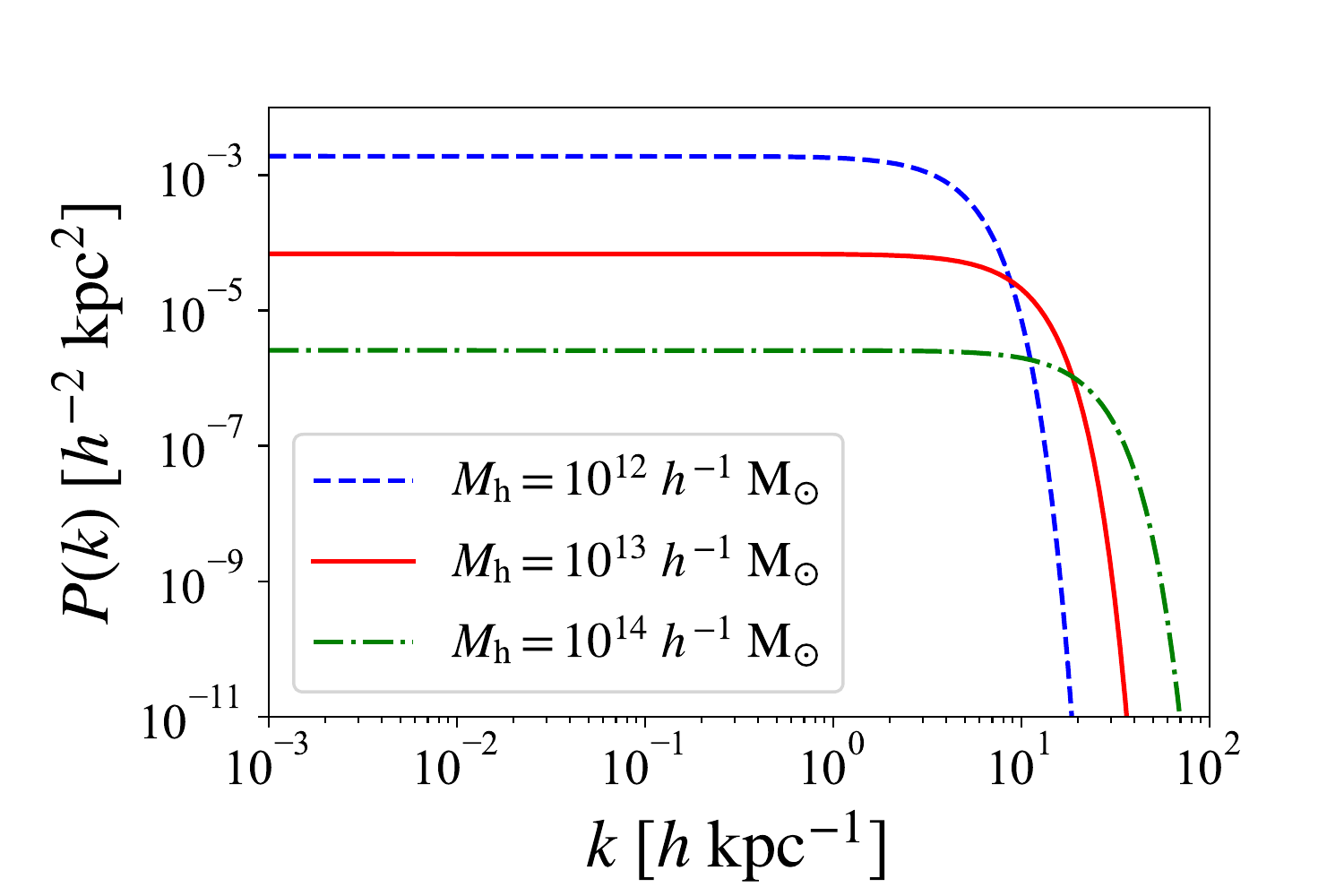}
 \caption{Halo mass dependence of the sub-galactic matter power spectrum. Two parameters $M_{\rm s}/M_{\rm h}$ and $m$ are fixed as $M_{\rm s}/M_{\rm h} = 0.01$ and $m = 10^{-22} \rm{eV}$, respectively. The position $x$ is set to one-hundredth of the virial radius of each halo, which is roughly the Einstein radius that we focus in Sec.~\ref{sec:compare-obs}. 
 For each case, we confirm that the radius $x$ is larger than the transition radius of the soliton core \citep{2014PhRvL.113z1302S}.}
 \label{Mhdep}
\end{figure}

\begin{figure}
 \centering
 \includegraphics[width=85mm]{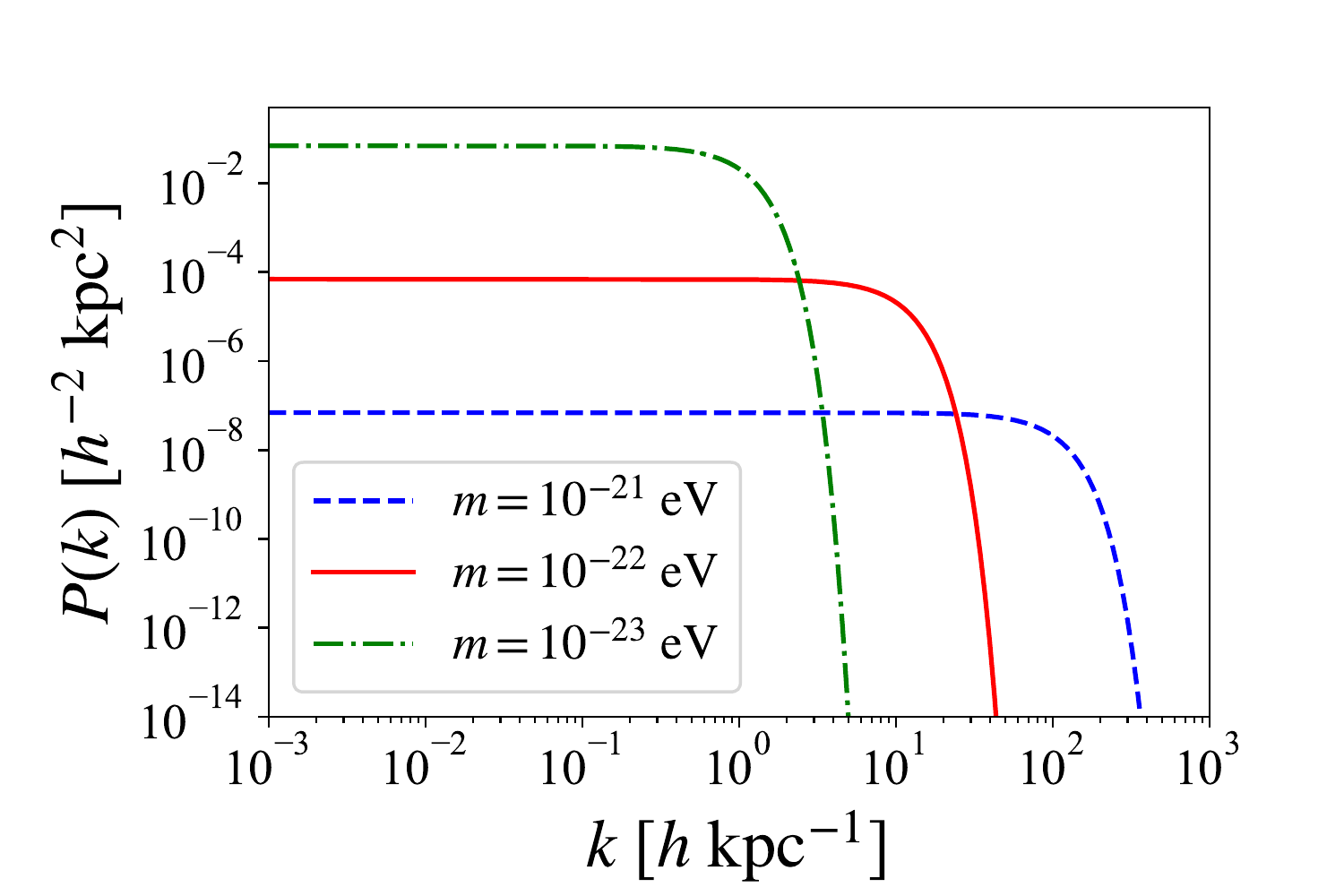}
 \caption{FDM mass dependence of the sub-galactic matter power spectrum. The other 3 parameters are fixed as $M_{\rm h} = 10^{13}\ h^{-1}\  M_{\rm{\odot}}, M_{\rm s}/M_{\rm h} = 0.01$, and $x$ being one-hundredth of the virial radius of the halo, $x \sim 4.4\ h^{-1}\ \rm{kpc}$.
 We note that $x$ is larger than the transition radii of the soliton core \citep{2014PhRvL.113z1302S}, which are $0.014$, $0.14$, and $1.4\ h^{-1}\ \rm{kpc}$ for FDM masses of $10^{-21}$, $10^{-22}$ and $10^{-23}\ \rm{eV}$, respectively.}
 \label{mdep}
\end{figure}

As mentioned in Sec.~\ref{subsec:fdmonly}, the soliton core is ignored in our model. 
Therefore our model is not valid around the central region in FDM halos. According to \citet{2014NatPh..10..496S}, the transition radius between the soliton core and NFW profile is around $3r_{\rm c}$, where $r_{\rm c}$  is the core radius that is known to scale with the FDM mass as well as the total halo mass \citep{2014PhRvL.113z1302S}.
Since we focus on Einstein radii in strong lens systems that are typically larger than the transit radius unless the FDM mass is too small (see also captions of Figs.~\ref{Mhdep} and \ref{mdep}), we can use our model to compare with strong lens observation as we attempt in Sec.~\ref{sec:compare-obs}. 
We also note that our model is likely invalid at around and beyond the virial radius as we do not include the dark matter distribution outside the virial radius.  

The normalized mass profile function $u(\boldsymbol{r}-\boldsymbol{r'})$ is assumed to be a spherical Gaussian function whose radial variance equals the half of the de Broglie wavelength. 
This assumption on the sphericity and Gaussian radial mass profile is consistent with finding in \citet{2021JCAP...03..076D} in which the FDM halo structure derived with a method proposed in \citet{1993ApJ...416L..71W} is found to be described well by a superposition of the randomly distributed spherical Gaussian clumps.
The Fourier transform of this function in the projected field is
\begin{equation}
    \widetilde{u}_{\boldsymbol{k}} = \widetilde{u}_{k} = \exp{\left(-\frac{\lambda_{\rm c}^{2}k^{2}}{8}\right)}. \label{massprofile}
\end{equation}

In order to calculate the de Broglie wavelength $\lambda_{\rm c} = 2 \pi \hbar/mv$, we set $v$ as
\begin{equation}
    v = \sqrt{\frac{3GM_{\rm tot}}{2R_{\rm vir}}}, \label{v}
\end{equation}
where $G$ is the gravitational constant, $M_{\rm tot}$ is the total mass that is the sum of the halo mass $M_{\rm h}$ and the stellar mass $M_{\rm s}$, and $R_{\rm vir}$ is the virial radius of the halo.
Note that $v$ is assumed to be constant within each halo throughout the paper.
An additional parameter to calculate $\lambda_{\rm c}$ is only FDM mass $m$.
Finally the position $x$ in the projected field is needed to calculate the power spectrum.

To sum up, there are 4 parameters, the total halo mass $M_{\rm h}$, the stellar-to-halo mass ratio $M_{\rm s}/M_{\rm h}$, the FDM mass $m$, and the position $x$ to calculate the sub-galactic matter power spectrum in FDM halos. 
Substituting the normalized mass profile function Eq.~(\ref{massprofile}), Eq.~(\ref{ps_wb}) becomes
\begin{equation}
    P(k) = \left(\frac{\Sigma_{\rm h}(x)}{\Sigma_{\rm h}(x) + \Sigma_{\rm b}(x)}\right)^{2} \frac{4\pi \lambda_{\rm c}^{3}}{3 r_{\rm h}(x)}\ \exp{\left(-\frac{\lambda_{\rm c}^{2}k^{2}}{4}\right)}. \label{ps_full} 
\end{equation}
With this model, we show the parameter dependence especially the total halo mass $M_{\rm h}$ and the FDM mass $m$.

Fig.~\ref{Mhdep} shows the total halo mass dependence. 
It is found that the power spectrum damps at the larger wavenumber with the larger total halo mass.
In addition, the amplitude of the plateau region is smaller with the larger total halo mass.
The former can be understood as follows. 
From Eq.~(\ref{ps_full}), we can find that the spectrum damps around $k \sim 1/\lambda_{\rm c}$. 
From Eq.~(\ref{v}), we have approximately $v \propto M_{\rm h}^{1/3}$ since we can approximate as $M_{\rm tot} \propto M_{\rm h}$ and $R_{\rm vir} \propto M_{\rm h}^{1/3}$.
The de Broglie wavelength scales as $\lambda_{\rm c} \propto M_{\rm h}^{-1/3}$.
Therefore the damping scale is different among different total halo masses, even if the FDM mass $m$ is fixed.
The latter result can be understood as follows.
In the plateau region, $P(k) \propto \lambda_{\rm c}^{3}/r_{\rm h}(x) \propto M_{\rm h}^{-4/3}$, if we approximate $r_{\rm h}(x) \propto R_{\rm vir}$. 

These results can also be understood qualitatively as follows.
The variance in real space is obtained by
\begin{equation}
    \sigma_{\delta}^{2} \sim \int d^{2}k\ P(k) \sim \mathcal{O}\left(\frac{\lambda_{\rm c}}{r_{\rm h}} \right) \sim \mathcal{O}\left(\frac{1}{N}\right),\label{variance}
\end{equation}
where $N \sim r_{\rm h}/\lambda_{\rm c}$ is the number of clumps along the line of sight.
From Eq.~(\ref{variance}), the fluctuation along the line of sight can be approximated as $\mathcal{O}(1/\sqrt{N})$, which is consistent with a naive picture that $\mathcal{O}(1)$ fluctuations of individual clumps are averaged out by $N$ clumps along the line of sight.
As the total halo mass becomes larger, the virial radius becomes larger and the number of quantum clumps along the line of sight increases.
The large number of clumps along the line of sight results in the smaller amplitude of the power spectrum due to averaging.
    
Fig.~\ref{mdep} shows the FDM mass dependence. 
It is found that the power spectrum damps at the larger wavenumber and the amplitude in the plateau region is smaller with the larger FDM mass.
Since the FDM mass and the de Broglie wavelength are related with each other by $\lambda_{\rm c} \propto m^{-1}$, the power spectrum in the plateau region is proportional to $P(k) \propto m^{-3}$, and the damping scale is $k \propto m$.
These results can also be understood in the same way as in the discussion above.
As the FDM mass becomes larger, the de Broglie wavelength becomes smaller, 
leading to the larger number of the clumps along the line of sight, and the lower amplitude of the power spectrum.
Since the sub-galactic matter power spectrum is sensitive to the FDM mass, it can be used to constrain the mass range of FDM.

In addition, Fig.~1 in \citet{2016JCAP...11..048H} shows that the sub-galactic power spectrum due to CDM subhalos is around $10^{-6}\  h^{-2}\ \rm{kpc}^{2}$ in the small wavenumber limit, which is much smaller than the sub-galactic matter power spectrum in the FDM model in most cases of interest (see also Chan et al. 2020).
It suggests that we can obtain interesting constraints on FDM mass around the typical range from observations of the sub-galactic matter power spectrum, which we discuss in Sec.~\ref{sec:compare-obs}. 

\section{Comparison with observation} \label{sec:compare-obs}
Using our formalism described in Sec.~\ref{sec:power-spectrum}, we compare the sub-galactic matter power spectrum with a real observational data to constrain the range of FDM mass.
We first use the current constraint on the sub-galactic matter power spectrum \citep{2018arXiv180305952B} that is obtained from the SLACS strong lens system SDSS J0252+0039 \citep{2009ApJ...705.1099A}. 
Next we discuss the future prospect of constraints that are also obtained by strong lens systems \citep{2016JCAP...11..048H}.
In Sec.~\ref{subsec:dimless-ps}, we define the dimensionless convergence power spectrum.
We show the comparison with the current data in Sec.~\ref{subsec:current-obs}, and the future prospect in Sec.~\ref{subsec:future-obs}.

\subsection{Dimensionless convergence power spectrum} \label{subsec:dimless-ps}
We use the dimensionless convergence power spectrum described below when we compare our model with the observation.
Consider a lens system with the angular diameter distance from the observer to the lens $D_{\rm d}$, from the observer to the source $D_{\rm s}$ and from the lens to the source $D_{\rm ds}$.
The critical surface-mass density $\Sigma_{\rm cr}$ for this lens system is given by
\begin{equation}
    \Sigma_{\rm cr} = \frac{1}{4\pi G} \frac{D_{\rm s}}{D_{\rm d} D_{\rm ds}}.
\end{equation}
With this critical surface-mass density and projected density field (Eq.~(\ref{Sigma_tot})), we can define a convergence field $\kappa(\boldsymbol{x})$ as
\begin{equation}
    \kappa(\boldsymbol{x}) = \frac{\Sigma(\boldsymbol{x})}{\Sigma_{\rm cr}}.
\end{equation}
The convergence power spectrum is defined as 
\begin{equation}
    \left<\widetilde{\kappa}_{\boldsymbol{k}}\ \widetilde{\kappa}_{\boldsymbol{k'}} \right> \equiv S_{\epsilon} \delta^{(2)}_{\boldsymbol{k}+\boldsymbol{k'}, \boldsymbol{0}}\ P_{\delta \kappa}(k), 
\end{equation}
where $\widetilde{\kappa}$ is the two dimensional Fourier transform of the convergence field.
Additionally we define the dimensionless convergence power spectrum, which we use for the comparison with the observation, as
\begin{equation}
    \Delta_{\delta \kappa}^{2}(k) = 2\pi k^{2} P_{\delta \kappa}(k).
\end{equation}
From this relation we can relate the convergence power spectrum and matter power spectrum in Sec.~\ref{sec:power-spectrum} as,
\begin{equation}
   \Delta_{\delta \kappa}^{2}(k) = 2\pi k^{2} \left(\frac{\Sigma(x)}{\Sigma_{\rm cr}}\right)^{2} P(k).
\end{equation}
We use this relation and compare with observational result with the dimensional convergence power spectrum.

\subsection{Current observation} \label{subsec:current-obs}
From Table~3 and Table~4 in \citet{2009ApJ...705.1099A}, we can obtain the information of SDSS J0252+0039.
Redshifts of the lens and the source are $z_{\rm lens} = 0.280$ and $z_{\rm src} = 0.982$, respectively, from which we obtain $\Sigma_{\rm cr} = 4.0 \times 10^{9}\ h\ M_{\odot}\ \rm kpc^{-1}$.
In order to calculate the dimensionless convergence power spectrum in our model and to constrain the FDM mass, we need to determine the NFW profile, the Hernquist profile and its position.
We adopt the Salpeter stellar mass $M^{\rm Salp}_{\rm s} = 2.0 \times 10^{11}\ h^{-1}\ M_{\odot}$ and the scale radius $a = 2.34\ h^{-1}\ \rm kpc$ for the Hernquist profile derived in \citet{2009ApJ...705.1099A}.
Note that the scale radius is converted from the scale radius $r_{\rm e}$ in $I$-band by $a = r_{\rm e}/1.8153$.

The NFW profile can be obtained as follows.
Since we know the stellar mass fraction within the Einstein radius $f_{\rm Ein}^{\rm Salp}= 0.71, r_{\rm Ein} = 3.1\ h^{-1}\ \rm kpc$ from the strong lensing observation, as well as the stellar mass distribution, we can determine the total mass of the NFW profile such that it recovers the observed stellar mass fraction.
We obtain the halo mass of $M_{\rm h} = 6.3\times 10^{12}\ h^{-1}\ M_{\odot} $.
The position is set to the observed Einstein radius, $x = 3.1\ h^{-1}\ \rm kpc $. 

Fig.~\ref{m_constraint} shows the result. 
The orange region shows the current excluded region of the sub-galactic power spectrum obtained by \citet{2018arXiv180305952B}, $\Delta^{2}_{\delta \kappa} < 1$ on 0.5 kpc, $\Delta^{2}_{\delta \kappa} < 0.1$ on 1 kpc and $\Delta^{2}_{\delta \kappa} < 0.01$ on 3 kpc at the 99 percent confidence level.
We find that no constraint of the FDM mass can be obtained.
This is because the current constraint by \citet{2018arXiv180305952B} is quite conservative and as a result is not tight enough. 

\subsection{Future constraint} \label{subsec:future-obs}
\begin{figure}
 \centering
 \includegraphics[width=85mm]{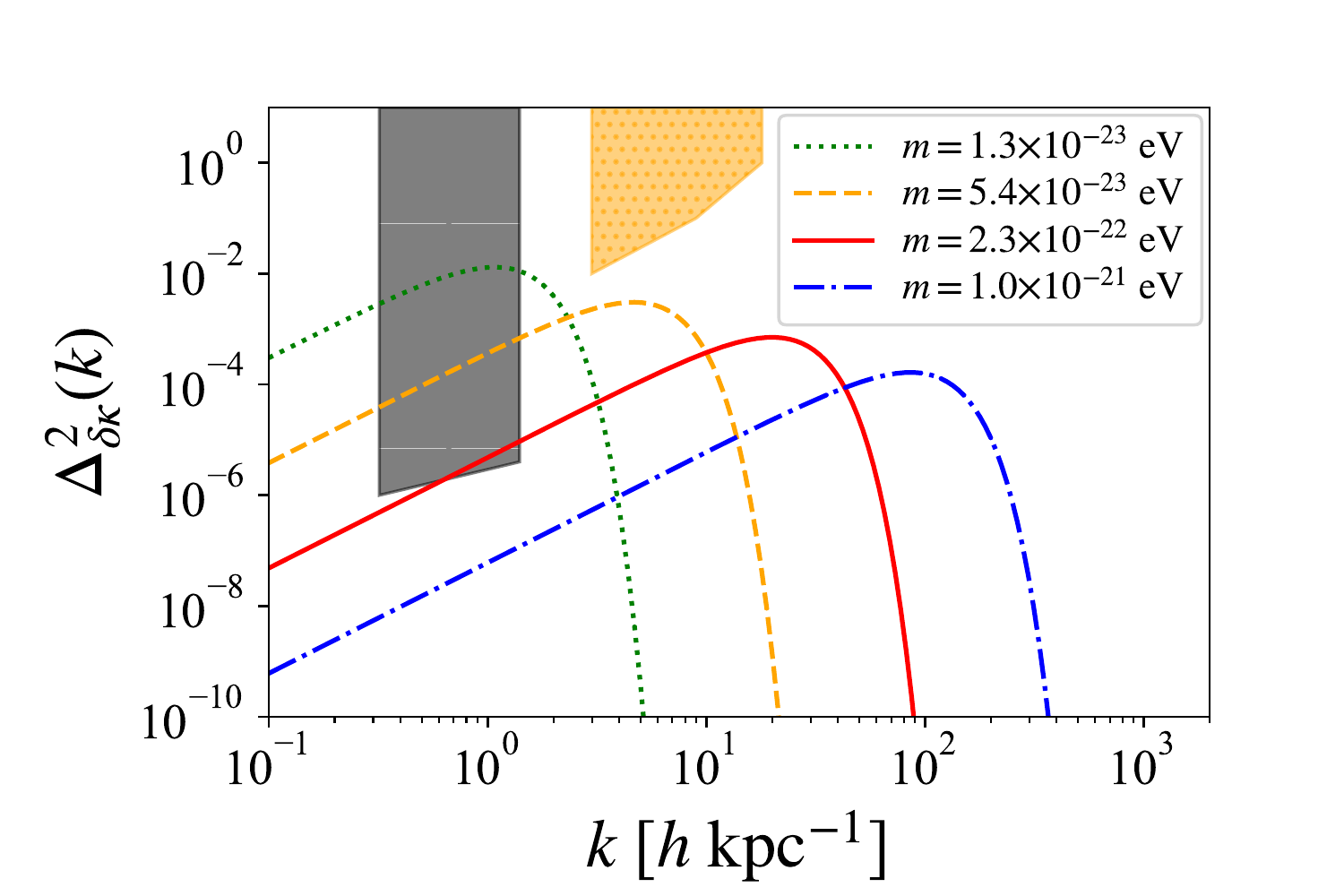}
 \caption{Comparison between our model of the dimensionless convergence power spectrum with the current and the future constraints. 
 The orange region is the currently excluded region obtained by \citet{2018arXiv180305952B} from the analysis of the strong lens system SDSS J0252+0039. 
 The black region is the expected excluded region obtained by future ALMA observations of strong lens systems \citep{2016JCAP...11..048H}. 
 We set parameters in our model calculation to those consistent with SDSS J0252+0039. 
 Specifically we use the NFW profile with halo mass $M_{\rm h} = 6.3\times 10^{12}\ h^{-1}\ M_{\odot}$, the Hernquist profile with stellar mass $M^{\rm Salp}_{\rm s} = 2.0 \times 10^{11}\ h^{-1}\ M_{\odot}$ and the scale radius $a = 2.34\ h^{-1}\ \rm kpc$, and the position is set to the observed Einstein radius, $x = 3.1\ h^{-1}\ \rm kpc$. 
 We vary the FDM mass in the range $m = \mathcal{O}(10^{-23}) - \mathcal{O}(10^{-21})\ \rm eV$.
 }
 \label{m_constraint}
\end{figure}

The black region in Fig.~\ref{m_constraint} shows the expected  excluded region of the sub-galactic matter power spectrum at the 68 percent confidence level obtained by future ALMA observations of strong lens systems \citep{2016JCAP...11..048H}.
We find that the FDM mass with $m \lesssim 3.2\times 10^{-22}\ \rm eV $ can be excluded by the future ALMA observation of SDSS J0252+0039-like strong lens systems, if no significant signal of the sub-galactic matter power spectrum is measured by such future observations.
Note that we can also obtain the lower bound of the FDM mass.

\begin{figure}
 \centering
 \includegraphics[width=85mm]{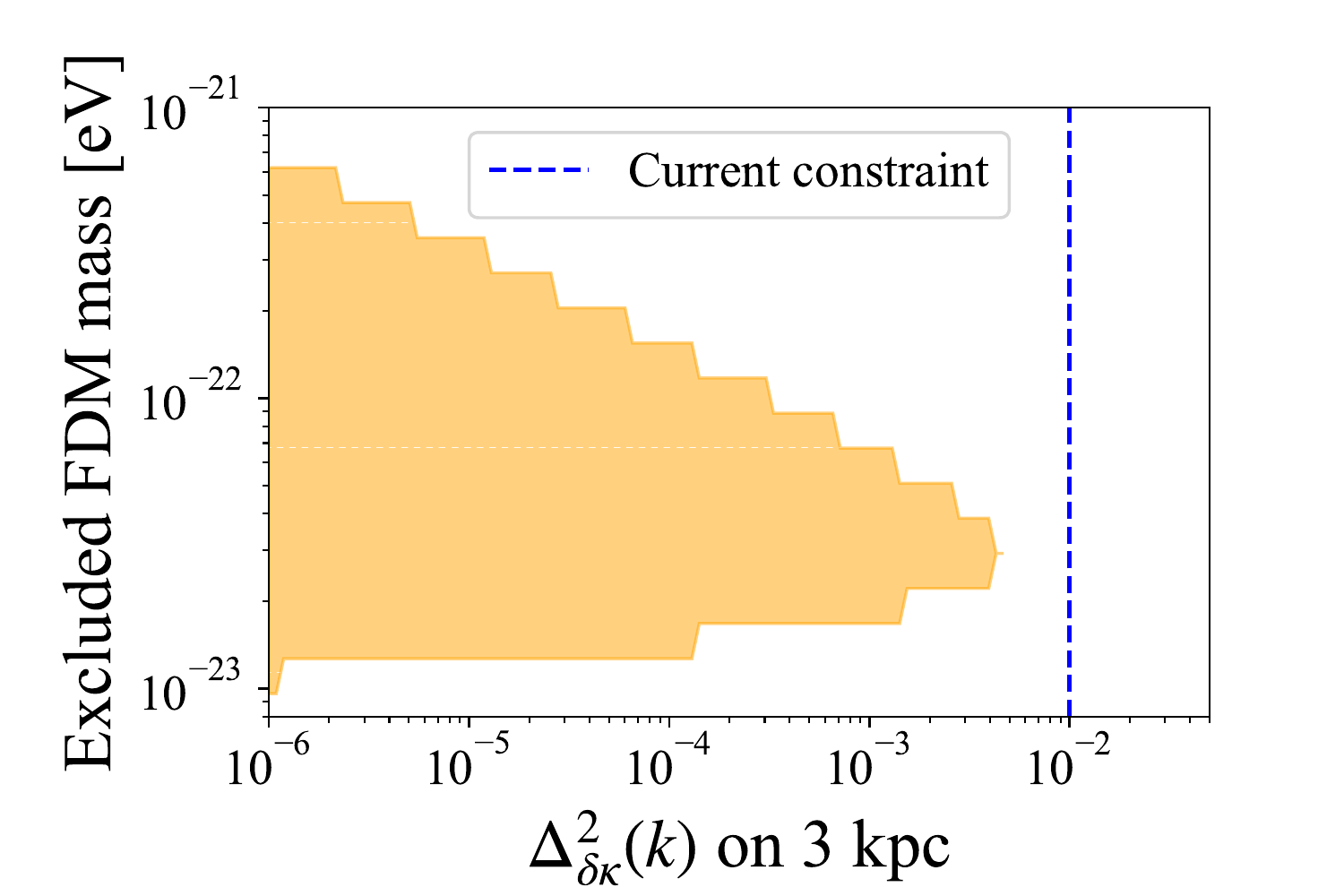}
 \caption{The excluded range of the FDM mass as a function of the future constraint on the dimensionless convergence power spectrum on 3 kpc, assuming observations of SDSS J0252+0039-like strong lens systems. 
 The parameters are the same as those used in Fig.~\ref{m_constraint}.}
 \label{m_future}
\end{figure}

We now discuss the relation between the excluded FDM mass range and the dimensionless convergence power spectrum probed by the future observation of SDSS J0252+0039-like strong lens systems.
Fig.~\ref{m_future} shows the relation between the excluded mass range and the future constraint on the dimensionless convergence power spectrum on 3 kpc.
It is found that the interesting mass range of the FDM can be constrained by future observations.
Here we show the constraint obtained from 3 kpc wavelength, although we can use other wavelengths as well.
Given the FDM mass dependence shown in Fig.~\ref{m_constraint}, we can constrain the lower mass if we use higher wavelengths or lower wavenumbers, and the higher mass if we use lower wavelengths or higher wavenumbers.

\section{Conclusion} \label{sec:conclusion}
In this paper, we first provide an analytic model for the sub-galactic matter power spectrum in FDM halos, assuming that the distribution of FDM in halos is described by a superposition of  quantum clumps whose size is comparable to the de Broglie wavelength of FDM.
We derive the power spectrum projected along the line of sight that can be directly compared with that measured from strong lens observations.
We show that this power spectrum has a large dependence on the FDM mass, suggesting that it is a useful probe of the FDM model.
In order to constrain the FDM mass, we compare our model with the current constraint from the galaxy-galaxy strong lens system SDSS J0252+0039. 
While the current constraint is not tight enough to give useful constraint on the FDM mass, we find that future observations can tightly constrain the FDM model in the interesting range of the FDM mass around $\mathcal{O}(10^{-22})\ \rm eV$.

When this paper was almost completed, we found a paper by \cite{2021arXiv210901168I} in which a new measurement of sub-galactic matter power spectrum for the strong lens system MG0414+0534 is presented. They measure the dimensionless convergence power spectrum of $\Delta^2_{\delta\kappa}\sim 5\times 10^{-4}$ on $\sim 8$~kpc scale and interpret the signal only in the context of the standard CDM model. We adopt lens model parameters of MG0414+0534 shown in \citet{2014MNRAS.439.2494O} to repeat the calculation in this paper, and find that quantum clumps in the FDM model with the mass of $m\approx 4\times 10^{-23}$~eV well explains the observed signal. However a caveat is that MG0414+0534 has been known as a quasar lens system exhibiting the significant flux ratio anomaly as well as strong perturbations on the lens potential due to a satellite galaxy  \citep[e.g.,][]{2009ApJ...697..610M}, which suggests that the measured sub-galactic matter power spectrum might be biased high. The high source redshift of $z=2.64$ also implies that the effect of the line-of-sight structure may be more important than for SDSS J0252+0039.

\begin{acknowledgments}
We thank the anonymous referee for useful comments.
We thank Naoki Yoshida for giving us useful advice.
This work was supported in part by World Premier International Research Center Initiative (WPI Initiative), MEXT, Japan, and JSPS KAKENHI Grant Number JP20H04725, JP20H00181, JP20H05856, JP18K03693.
J. Lim acknowledges support from the Research Grants Council of Hong Kong through the GRF grant 17304519, which also provides partial support for the postgraduate fellowship of A. Alfred.
\end{acknowledgments}

\bibliography{ref}
\bibliographystyle{aasjournal}

\end{document}